\documentclass[twocolumn,showpacs,preprintnumbers,amsmath,amssymb]{revtex4}
\usepackage{colordvi}
\usepackage{color}
\usepackage{epsfig}
\usepackage{graphicx}
\usepackage{bm}

\def\a {\not{\!\epsilon}}
\def\q {\not{\!k}}

\def\M  {{\cal M}}
\def\F  {{\cal F}}
\def\O  {{\cal O}}
\def\P  {{\cal P}}
\def\R  {{\cal R}}
\def\pc {\not{\!p_c}}
\def\pb {\not{\!p_{\bar{c}}}}
\def\k  {\mbox{\bf k}}
\def\p  {\mbox{\bf p}}
\def\Q  {\mbox{\bf Q}}
\def\e  {$\eta_c$ }
\def\J  {$J/\psi$ }
\def\cc {$c\bar{c}^{[8]}$ }
\def\ktf {$k_T$-factorization}
\def\ktfa {$k_T$-factorization approach}
\def\cpc#1#2#3  {{Chin.\ Phys.\ C }          {\bf#1}, #2 (#3)}
\def\err#1#2#3  {{\it Erratum }              {\bf#1}, #2 (#3)}
\def\epjc#1#2#3 {{Eur.\ Phys.\ J.\ C }       {\bf#1}, #2 (#3)}
\def\dum#1#2#3  {{~}                         {\bf#1}, #2 (#3)}
\def\ib#1#2#3   {{\it ibid. }                {\bf#1}, #2 (#3)}
\def\jcp#1#2#3  {{J.\ Comp.\ Phys.\ }        {\bf#1}, #2 (#3)}
\def\jhep#1#2#3 {{JHEP }                     {\bf#1}, #2 (#3)}
\def\ijmp#1#2#3 {{Int.\ J.\ Mod.\ Phys.\ }   {\bf#1}, #2 (#3)}
\def\jpg#1#2#3  {{J.\ Phys.\ G }             {\bf#1}, #2 (#3)}
\def\mpla#1#2#3 {{Mod.\ Phys.\ Lett.\ A }    {\bf#1}, #2 (#3)}
\def\ncim#1#2#3 {{Nuovo Cimento }            {\bf#1}, #2 (#3)}
\def\np#1#2#3   {{Nucl.\ Phys.\ }            {\bf#1}, #2 (#3)}
\def\npb#1#2#3  {{Nucl.\ Phys.\ B }          {\bf#1}, #2 (#3)}
\def\pan#1#2#3  {{Phys.\ At.\ Nuclei }       {\bf#1}, #2 (#3)}
\def\plb#1#2#3  {{Phys.\ Lett.\ B }          {\bf#1}, #2 (#3)}
\def\prep#1#2#3 {{Phys.\ Rep.\ }             {\bf#1}, #2 (#3)}
\def\prd#1#2#3  {{Phys.\ Rev.\ D }           {\bf#1}, #2 (#3)}
\def\prl#1#2#3  {{Phys.\ Rev.\ Lett.\ }      {\bf#1}, #2 (#3)}
\def\ptp#1#2#3  {{Prog.\ Theor.\ Phys.\ }    {\bf#1}, #2 (#3)}
\def\ps#1#2#3   {{Physica Scripta }          {\bf#1}, #2 (#3)}
\def\rmp#1#2#3  {{Rev.\ Mod.\ Phys.\ }       {\bf#1}, #2 (#3)}
\def\rpp#1#2#3  {{Rep.\ Prog.\ Phys.\ }      {\bf#1}, #2 (#3)}
\def\sa#1#2#3   {{Sci.\ Acta }               {\bf#1}, #2 (#3)}
\def\sjnp#1#2#3 {{Sov.\ J.\ Nucl.\ Phys.\ }  {\bf#1}, #2 (#3)}
\def\spj#1#2#3  {{Sov.\ Phys.\ JETP }        {\bf#1}, #2 (#3)}
\def\spjl#1#2#3 {{Sov.\ JETP Lett.\ }        {\bf#1}, #2 (#3)}
\def\spu#1#2#3  {{Sov.\ Phys.-Usp.\ }        {\bf#1}, #2 (#3)}
\def\yaf#1#2#3  {{Yad.\ Fiz.\ }              {\bf#1}, #2 (#3)}
\def\zp#1#2#3   {{Zeit.\ Phys.\ }            {\bf#1}, #2 (#3)}
\def\zpa#1#2#3  {{Z.\ Phys.\ A }             {\bf#1}, #2 (#3)}
\def\zpc#1#2#3  {{Z.\ Phys.\ C }             {\bf#1}, #2 (#3)}
\def\et {{\it et al.}}
\newcommand{\n}{\nonumber \\}
\begin{document}
\draft
\title{Prompt $\eta_c$ meson production at the LHC in the NRQCD with \ktf}
\author{S.\ P.\ Baranov}
\email{baranovsp@lebedev.ru}
\affiliation{P.N. Lebedev Institute of Physics, 
              Lenin Avenue 53, 119991 Moscow, Russia}
\author{A.\ V.\ Lipatov}
\email{lipatov@theory.sinp.msu.ru}
\affiliation{Skobeltsyn Institute of Nuclear Physics,
      Lomonosov Moscow State University, 119991 Moscow, Russia}
\affiliation{Joint Institute for Nuclear Research, Dubna 141980, Moscow region, Russia}
      
%

\date{\today}
\begin{abstract}
In the framework of the \ktfa, the prompt production of \e mesons at the LHC
conditions is studied. Our consideration is based on the off-shell amplitudes
for hard partonic subprocesses and on the nonrelativistic QCD (NRQCD) formalism 
for the formation of bound states. We try two latest parametrizations for 
noncollinear, or transverse momentum dependent (TMD) gluon densities derived from 
the Catani-Ciafaloni-Fiorani-Marchesini (CCFM) equation. We use the values of the 
nonperturbative matrix elements obtained from a combined fit of the \e and
\J differential cross sections. Finally, we show an universal set of parameters
that provides a reasonable simultaneous description for all of the available 
data on the prompt \J and \e production at the LHC.

\end{abstract}
\maketitle

\section{Motivation}

Since long ago, the production of quarkonium states in high energy hadronic
collisions remains an area of intense attention from both theoretical and
experimental sides.
Our present work continues the line started in the previous publications
\cite{part1,part2,part3}. We have already considered there the production of
$\psi'$, $\chi_c$, and $J/\psi$ mesons and now come to $\eta_c$ mesons.
As usual, we work in the \ktfa. 

It is worth mentioning that the case of $\eta_c$ mesons turned out to be
rather puzzling for conventional NRQCD calculations at next-to-leading 
order (NLO) \cite{Kniehl,HanMa}. This time, the theory was very unlucky 
to have too few free adjustable parameters. Having the nonperturbative matrix 
elements (NMEs) fixed from fitting all other production data, the theory lost 
its flexibility and made a prediction for $\eta_c$ by a huge factor off the 
measured cross section. The overall situation was even called `challenging' 
\cite{Kniehl}. The aim of the present note is to show that the approach used
consistently in \cite{part1,part2,part3} meets no troubles with the \e data.

\section{Theoretical framework}

As it was done previously for $\psi'$, $\chi_c$ and $J/\psi$ production 
\cite{part1,part2,part3}, the present calculations are based on perturbative 
QCD and nonrelativistic bound state formalism (NRQCD).
The production of \e mesons is dominated by the color singlet (CS) contribution
that refers to the partonic subprocess
\begin{equation} g^*(k_1)+g^*(k_2)\to\eta_c(p) \end{equation}
with the respective cross section
\begin{eqnarray}
&&\sigma(pp\to\eta_c+X)\n
&&=\int \frac{2\pi}{x_1 x_2 s\,F}\,
{\F}_g(x_1,{\k}_{1T}^2,\mu^2)\,{\F}_g(x_2,{\k}_{2T}^2,\mu^2)\n
&&\times\; \bigl|{\cal M}(g^*g^*\to\eta_c)\bigr|^2\, \label{lips}
d{\k}_{1T}^2\,d{\k}_{2T}^2\,dy_\eta\,\frac{d\phi_1}{2\pi}\,\frac{d\phi_2}{2\pi},
\end{eqnarray}
where ${k}_1$ and ${k}_2$ denote the initial gluon 4-momenta, 
$\phi_1$ and $\phi_2$ are the respective azimuthal angles,
$y_\eta$ is the rapidity of \e meson,
$x_1$ and $x_2$ are the gluon longitudinal momentum fractions,
${\cal M}(g^*g^*\to\eta_c)$ is the hard scattering amplitude,
and ${\F}_g(x_i,{\k}_{iT}^2,\mu^2)$ is the transverse momentum dependent 
(TMD, or unintegrated) gluon density in a proton.
In accordance with the general definition \cite{BycKaj}, the off-shell gluon 
flux factor in (\ref{lips}) is taken as $F=2\lambda^{1/2}(\hat{s},k_1^2,k_2^2)$,
where $\hat{s}=(k_1+k_2)^2$. 

In addition to the above, we have considered a number of color octet (CO) contributions
and contribution from the feed-down $h_c\to\eta_c\,X$ process. The CO terms refer to the perturbative production of a color-octet 
\cc pair followed by nonperturbative gluon radiation bringing the intermediate 
\cc state to a real (colorless) meson: 
\begin{equation} 
g^*(k_1)+g^*(k_2)\to c\bar{c}^{[8]}\to\eta_c(p) + \mbox{soft~gluons}.
\end{equation}
The intermediate color octet \cc state can be either of
$^1S_0$, $^3S_1$, $^3P_0$, $^3P_1$, $^3P_2$, or $^1P_1$, where we use
standard spectroscopic notation.
The probabilities of the subsequent nonperturbative soft transitions are not
calculable within the theory and are usually accepted as free model parameters.
There are, however, certain restrictions coming from some general principles.
Whenever calculable or not, the nonperturbative amplitudes must be identical
for transitions in both directions (i.e., from vectors to scalars and vice
versa), as it is motivated by the heavy quark spin symmetry (HQSS). The
amplitudes can only differ by an overall normalizing factor representing the 
averaging over spin degrees of freedom. Thus, we strictly have from this
property \cite{Bodwin}:
\begin{eqnarray}
\left\langle \O^{\eta_c}\bigl[^1S_0^{[1]}\bigr]\right\rangle &=& \frac{1}{3}
\left\langle \O^{J/\psi}\bigl[^3S_1^{[1]}\bigr]\right\rangle \n
\left\langle \O^{\eta_c}\bigl[^1S_0^{[8]}\bigr]\right\rangle &=& \frac{1}{3}
\left\langle \O^{J/\psi}\bigl[^3S_1^{[8]}\bigr]\right\rangle \n
\left\langle \O^{\eta_c}\bigl[^3S_1^{[8]}\bigr]\right\rangle &=&
\left\langle \O^{J/\psi}\bigl[^1S_0^{[8]}\bigr]\right\rangle \n
\left\langle \O^{\eta_c}\bigl[^1P_1^{[8]}\bigr]\right\rangle &=& 3
\left\langle \O^{J/\psi}\bigl[^3P_0^{[8]}\bigr]\right\rangle \n
\left\langle \O^{h_c}      \bigl[^1P_1^{[1]}\bigr]\right\rangle &=& 3
\left\langle \O^{\chi_{c0}}\bigl[^3P_0^{[1]}\bigr]\right\rangle \n
%
%
\left\langle \O^{h_c}      \bigl[^1S_0^{[8]}\bigr]\right\rangle &=& 3
\left\langle \O^{\chi_{c0}}\bigl[^3S_1^{[8]}\bigr]\right\rangle 
\end{eqnarray}
The above relations require a simultaneous fit for the \e and \J production
data. This fit turned out to be impossible in the traditional NRQCD scheme.
The calculated cross sections were either found to be at odds with the 
measurements \cite{Kniehl} or at odds with theoretical principles \cite{HanMa}.

The crucial point in the above papers is the presence of a large unwanted
contribution to the $\eta_c$ production cross section from the intermediate 
$^3S_1^{[8]}$ state (unwanted, as the \e production cross section is saturated 
by the color singlet channel alone; a fact, already pointed out in \cite{Likhoded}).
The corresponding nonperturbative matrix element is an HQSS counterpart of 
the $^1S_0^{[8]}$ matrix element engaged in the production of \J mesons, where 
it is needed to make the outgoing \J meson unpolarised: this spinless state is
employed to dilute strong \J polarization in other channels. Note by the way 
that the size of $\langle\O^{J/\psi}[^1S_0^{[8]}]\rangle$ matrix element used 
in \cite{Kniehl} is in conflict with the NRQCD quark relative velocity counting
rules. 

In our present approach, we follow the interpretation of nonperturbative
color octet transitions in terms of multipole radiation theory. 
Then, the final state \J mesons come nearly unpolarized \cite{E1}, either 
because of the cancellation between the $^3P_1^{[8]}$ and $^3P_2^{[8]}$ 
contributions, or as a result of two successive color-electric (E1) dipole 
transitions in the chain $^3S_1^{[8]}\to {^3\!P}_{J}^{[8]}\to J/\psi$ with 
$J = 0, 1, 2$. Thus, we can avoid the $^1S_0^{[8]}$ contribution to \J and, 
as a consequence, get rid of the $^3S_1^{[8]}$ contribution to \e production.

In the numerical analysis shown below, we tried two latest sets of 
TMD gluon densities in a proton, 
referred to as JH'2013 set 1 and JH'2013 set 2 \cite{JH2013}.
These gluon densities were obtained from CCFM evolution equation where the input
parametrization (used as boundary conditions) was fitted to the proton structure 
function $F_2(x,Q^2)$. 
Following \cite{PDG}, we take the charmonia masses $m(\eta_c) = 2.9839$~GeV, 
$m(h_c) = 3.52538$~GeV, $m(J/\psi) = 3.0969$~GeV and the branching fractions 
$B(J/\psi \to \mu^+ \mu^-) = 0.05961$ and $B(h_c \to \eta_c \gamma) = 0.51$.
The renormalization and factorization scales 
were set to $\mu_R^2=m^2+{\p}_T^2$ and $\mu_F^2=\hat{s}+{\Q}_T^2$, where $m$ 
and ${\p}_T$ are the mass and transverse momentum of the produced charmonium,
and ${\Q}_T$ is the transverse momentum of the initial off-shell gluon pair.
The choice of $\mu_R$ is rather standard for charmonia production, while
the unusual choice of $\mu_F$ is connected with the CCFM evolution (see
\cite{JH2013} for details).
The analytic expressions for the hard scattering amplitudes in (1) and (3) 
were otained using the algebraic manipulation system \textsc{form} \cite{FORM}.
The multidimensional phase space integration has been performed by means
of the Monte-Carlo technique using the routine \textsc{vegas} \cite{VEGAS}.

\section{Numerical results}

\begin{center}
\begin{table*}[t]\label{Tab_Jpsi}
\caption{Sets of NME's for \J production as determined from the different fits}
\begin{tabular}{lcccc}
\hline \hline
  & JH set 1 & JH set 2 &~Kniehl \et ~\cite{BK}~&~Gong \et ~\cite{GW}\\[1mm]
\hline
$\left\langle \O^{J/\psi}\bigl[^3S_1^{[1]}\bigr]\right\rangle/\mbox{GeV}^3$
  & 1.16    &  1.16     & 1.32               &  1.16  \\
$\left\langle \O^{J/\psi}\bigl[^1S_0^{[8]}\bigr]\right\rangle/\mbox{GeV}^3$
  &  0.0    &  0.0      & 0.304              &  0.097 \\
$\left\langle \O^{J/\psi}\bigl[^3S_1^{[8]}\bigr]\right\rangle/\mbox{GeV}^3$
  & $(4.2 \pm 0.9) \cdot 10^{-4}$ & $(1.6 \pm 0.2) \cdot 10^{-3}$ &~0.00168~&~-0.0046\\
$\left\langle \O^{J/\psi}\bigl[^3P_0^{[8]}\bigr]\right\rangle/\mbox{GeV}^5$
  & $0.023 \pm 0.002$   & $0.024 \pm 0.002$  &-0.00908 & -0.0214\\
\hline \hline
\end{tabular}
\end{table*}
\end{center}

\begin{figure}
\begin{center}
\epsfig{figure=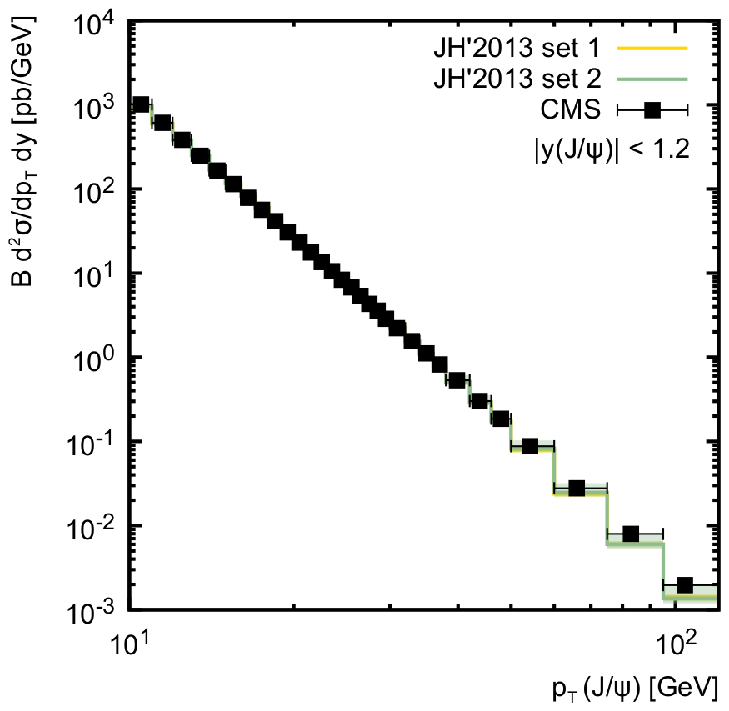, width = 4.2cm} 
\epsfig{figure=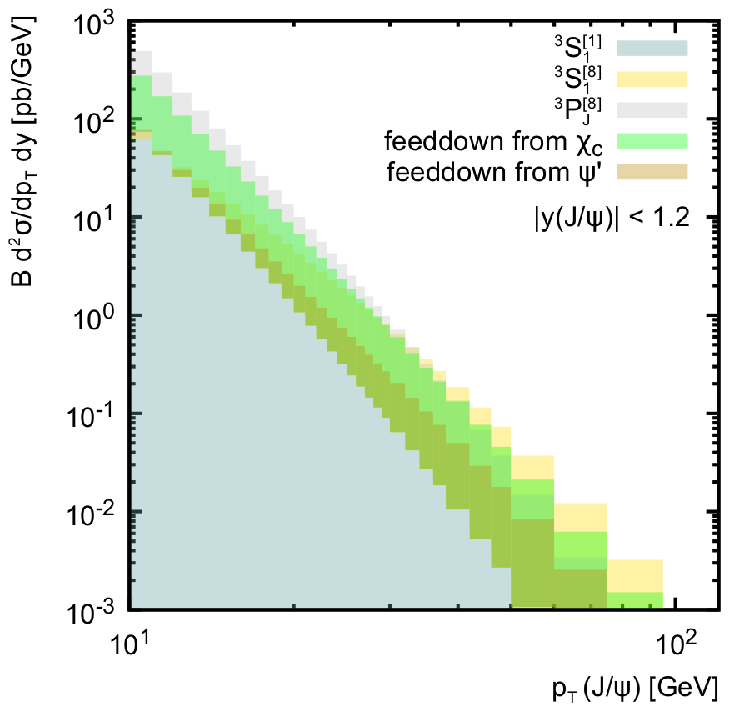, width = 4.2cm}
\epsfig{figure=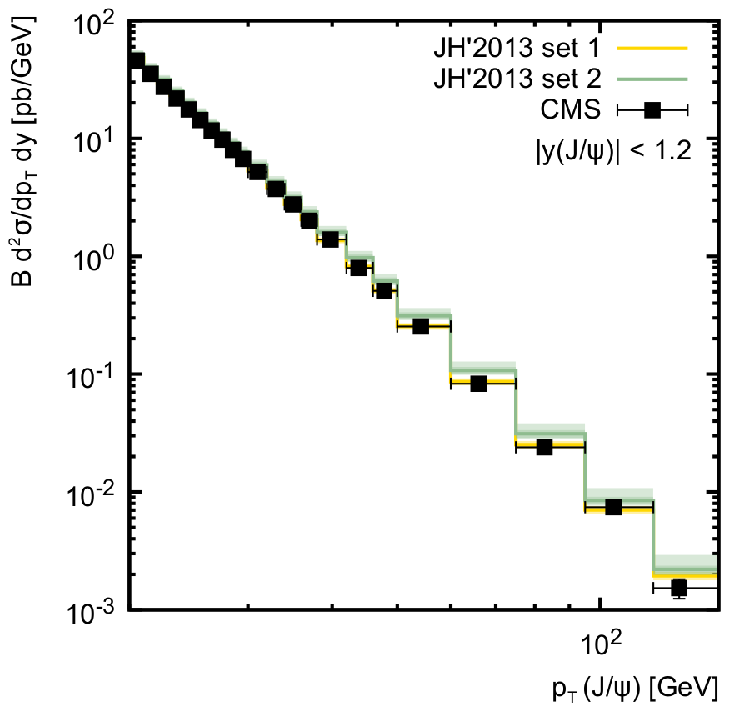, width = 4.2cm} 
\epsfig{figure=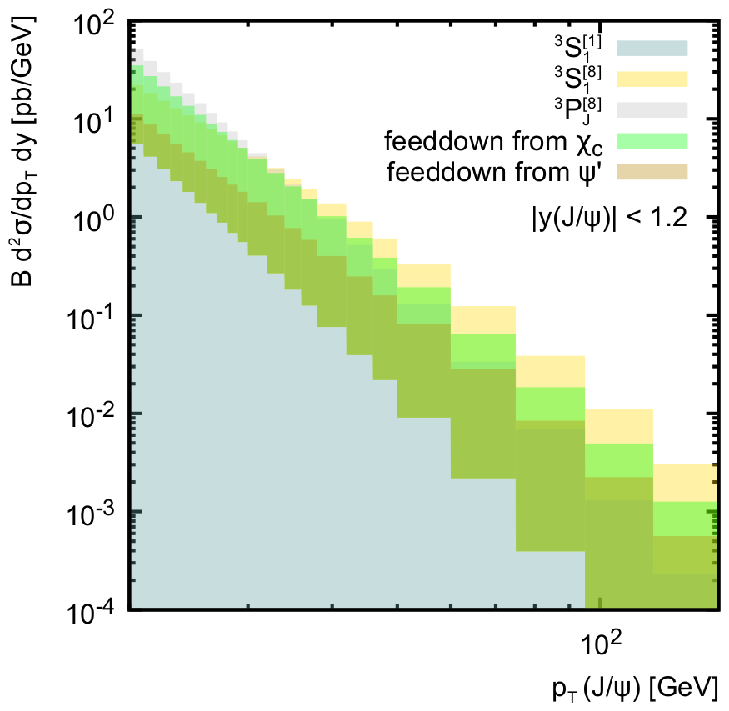, width = 4.2cm}
\caption{Transverse momentum distribution of prompt $J/\psi$ mesons
produced in $pp$ collisions at $\sqrt s = 7$~TeV (upper plots) and
$\sqrt s = 13$~TeV (lower plots). The shaded bands on the left panels represent
the total uncertainties of our calculations (i.e. scale uncertainties 
and the uncertainties coming from NMEs fit, summed in quadrature), as estimated 
for JH'2013 set 2 gluon density. The relative contributions from the different 
production mechanisms are shown on the right panels.
The experimental data are from CMS \cite{CMS}. }
\label{fig1}
\end{center}
\end{figure}

To determine the NMEs of \J mesons (as well as their \e counterparts)
we performed a combined fit of \J and \e transverse momentum distributions
using the latest CMS \cite{CMS}, ATLAS \cite{ATLAS} and LHCb data \cite{LHCb} 
collected at 7, 8 and 13 TeV. Here, the factorization
principle seems to be on solid theoretical grounds because of not too low $p_T$
values for both $J/\psi$ and $\eta_c$ mesons.
We do not impose any kinematic restrictions but the experimental acceptance.
The fitting procedure was separately done in each of the rapidity subdivisions
under the requirement that the NMEs be strictly positive, and then the 
mean-square average of the fitted values was taken.
Note that we used the results of a global fit for the entire charmonium family
(including, in particular, $\chi_{cJ}$ and $\psi'$ states) \cite{global} to
properly calculate the feed-down contributions from $h_c$, $\chi_{cJ}$, and 
$\psi'$ decays.


For some (yet unrecognized) reasons, our $^1P_1^{[1]}$ production amplitude 
(needed to calculate the feed-down $h_c\to\eta_c\,X$) disagrees with the one
found in the literature. 
Our calculation is off-shell, but has continuous on-shell limit that can
be promptly compared with \cite{Meijer,Troost}.
The contribution is anyway small and unimportant numerically; 
but the discrepancy is still of interest from the academic point of view. For 
the lack of details presented in \cite{Meijer,Troost}, we cannot repeat their 
calculation. The details of our calculation are explained in the Appendix.
\begin{figure}
\begin{center}
\epsfig{figure=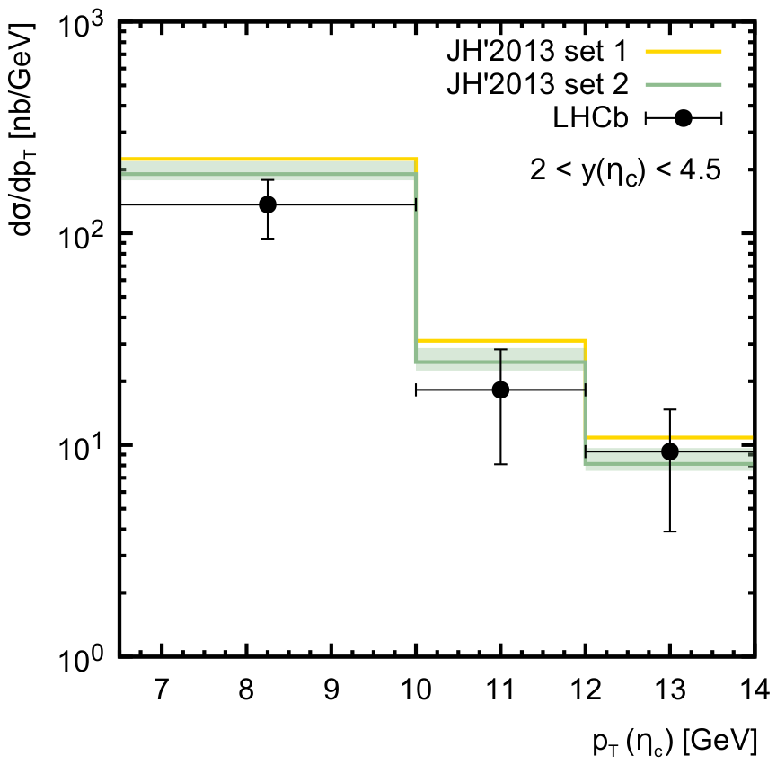, width = 4.2cm} 
\epsfig{figure=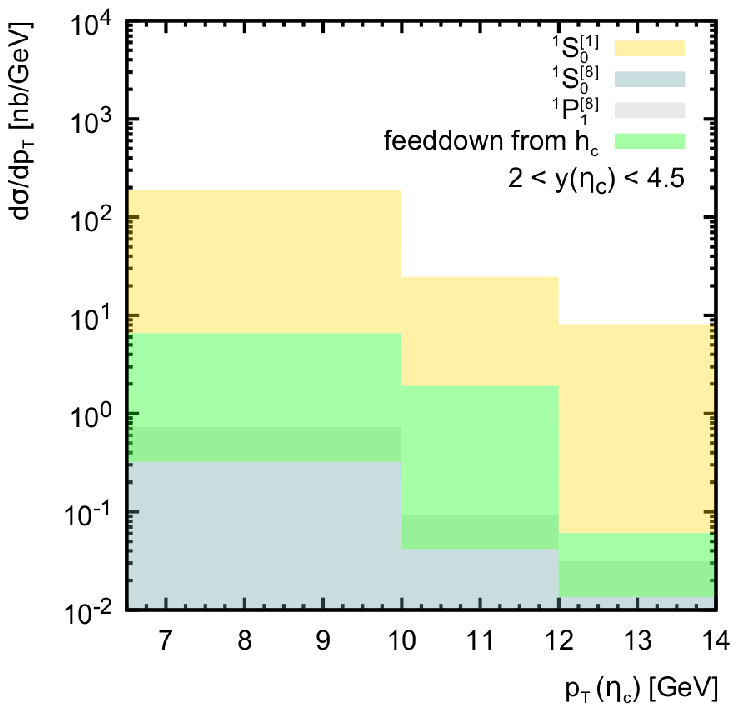, width = 4.2cm}
\epsfig{figure=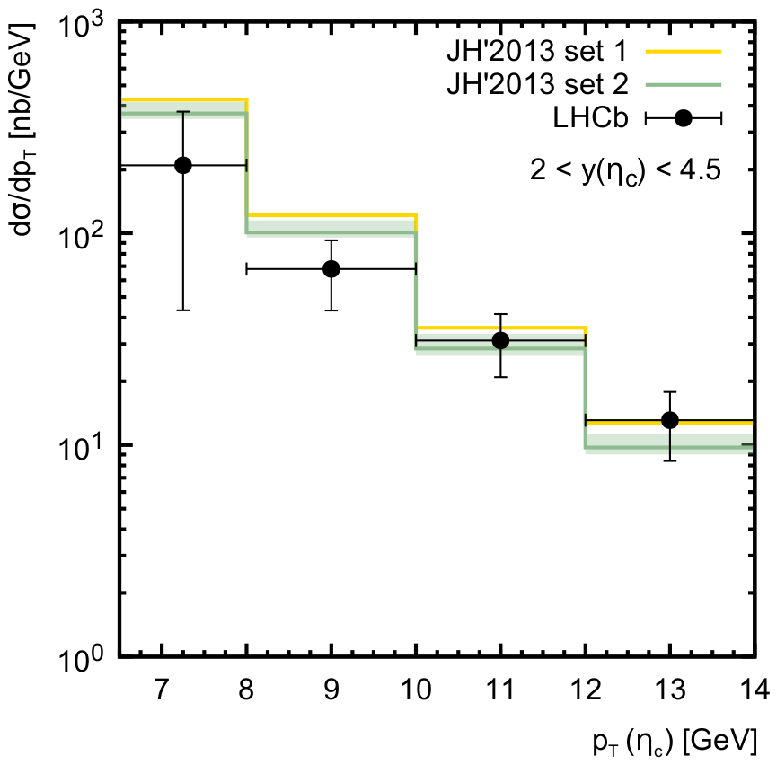, width = 4.2cm} 
\epsfig{figure=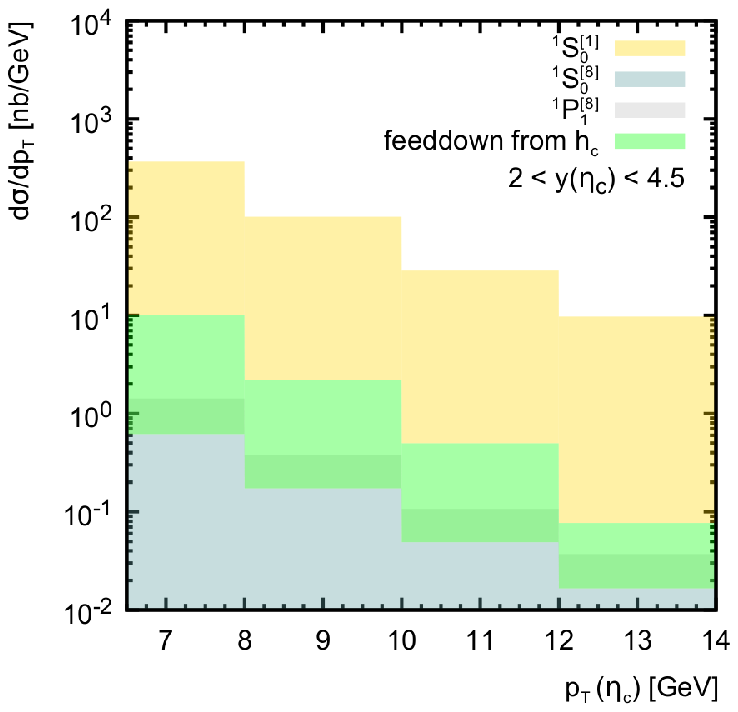, width = 4.2cm}
\caption{Transverse momentum distribution of prompt $\eta_c$ mesons 
produced in $pp$ collisions at $\sqrt s = 7$~TeV (upper plots) and
$\sqrt s =8$~TeV (lower plots). Shaded bands on the left panels represent
the total uncertainties of our calculations (i.e. scale uncertainties 
and the uncertainties coming from NMEs fit, summed in quadrature), as estimated
for JH'2013 set 2 gluon density. The relative contributions from the different 
production mechanisms are shown on the right panels.
The experimental data are from LHCb \cite{LHCb}. }
\label{fig2}
\end{center}
\end{figure}

The numerical values of our NMEs for $J/\psi$ and $h_c$ mesons are written out in Tables I and II.
For comparison, we also present here several sets of NMEs \cite{BK,GW, chic1, chic2}, obtained
in the NLO NRQCD by other authors. The NMEs shown for $h_c$ mesons are translated from 
$\chi_c$ NMEs using HQSS formulas. The fits differ from one another by somehow
differently selected data sets.
The corresponding values of NMEs for $\eta_c$ meson are collected in Table III.
They can be easily obtained 
from Table I using  the HQSS relations~(4).

A comparison of our predictions with the experimental results is displayed in 
Figs.~1 and~2. The theoretical uncertainty bands include both scale uncertainties 
and the uncertainties coming from the NMEs fitting procedure. First of them
were obtained by varying the $\mu_R$ scale around its default value by a factor 
of $2$. This was accompanied with using the JH'2013 set 2+ and JH'2013 set 2- 
in place of the JH'2013 set 2, in accordance with \cite{JH2013}. 
One can see that we have achieved a reasonably good agreement between our 
calculations and LHCb measurements (with both of the considered TMD gluons), 
simultaneously for the prompt $\eta_c$ and $J/\psi$ production data collected 
at different energies and in the whole $p_T$ range.
The presented results can give a significant impact on the understanding of 
charmonia production within NRQCD.

\begin{center}
\begin{table*}[t]\label{Tab_hc}
\caption{Sets of NME's for $h_c$ production as determined from the different fits}
\begin{tabular}{lcccc}
\hline \hline
  & JH set 1 & JH set 2 &~Zhang \et ~\cite{chic1}~&~Likhoded \et ~\cite{chic2}\\[1mm]
\hline
$\left\langle \O^{h_c}\bigl[^1P_1^{[1]}\bigr]\right\rangle/\mbox{GeV}^5$
  & $3.1 \pm 0.4$   &  $3.2 \pm 0.5$     & 0.96               &  4.51  \\
$\left\langle \O^{h_c}\bigl[^1S_0^{[8]}\bigr]\right\rangle/\mbox{GeV}^3$
  & $(6.0 \pm 3.0) \cdot 10^{-4}$   &  $(1.5 \pm 0.9) \cdot 10^{-3}$      & 0.00603             &  0.00132 \\
\hline \hline
\end{tabular}
\end{table*}
\end{center}

\section{Conclusions}

We have considered the production of charmonium states at the LHC and found 
a consistent simultaneous description for the \J and \e data. Our nonperturbative
matrix elements strictly obey the heavy quark spin symmetry rules.

The fundamental difference with the traditional NRQCD scheme (which was 
unable to accommodate the whole data set) is in a different treatement of the
nonperturbative color-octet transitions. The latter are interpreted in our
approach in terms of multipole radiation theory. Then the \J mesons are
produced unpolarized, thus making no need in a diluting $^1S_0^{[8]}$ 
contribution to \J production and, as a consequence, requiring no $^3S_1^{[8]}$ 
contribution to \e production.
In the forthcoming paper \cite{global} we are going to present a global fit
for the entire charmonium family, including $J/\psi$, $\chi_{cJ}$, $\psi(2S)$ 
and \e mesons.

\section{Appendix. Off-shell production amplitude for $^1P_1^{[1]}$ state}

In this section, we consider the gluon-gluon fusion subprocess
\begin{equation}\label{ggf}
g(k_1,\epsilon_1,a) + g(k_2,\epsilon_2,b)\to 
g(k_3,\epsilon_3,c) + c\bar c (p,\epsilon^\alpha),
\end{equation}
where the symbols in the parentheses indicate the momentum, the polarization, and 
the color of the interacting quanta. The calculation of this subprocess at
${\O}(\alpha_s^3)$ relates to six Feynman diagrams:
\begin{eqnarray}
{\M}_1&=&
 \mbox{tr}\!\left\{\a_1(\pc-\q_1+m_c)\a_2(-\pb-\q_3+m_c)\a_3\;{\P}_S\right\}\n
&\times&
 \left[k_1^2-2(p_{c}k_1)\right]^{-1}\left[k_3^2+2(p_{\bar{c}}k_3)\right]^{-1},\\
{\M}_2&=&
 \mbox{tr}\!\left\{\a_1(\pc-\q_1+m_c)\a_3(-\pb+\q_2+m_c)\a_2\;{\P}_S\right\}\n
&\times&
 \left[k_1^2-2(p_{c}k_1)\right]^{-1}\left[k_2^2-2(p_{\bar{c}}k_2)\right]^{-1},\\
{\M}_3&=&
 \mbox{tr}\!\left\{\a_3(\pc+\q_3+m_c)\a_1(-\pb+\q_2+m_c)\a_2\;{\P}_S\right\}\n
&\times&
 \left[k_3^2+2(p_{c}k_3)\right]^{-1}\left[k_2^2-2(p_{\bar{c}}k_2)\right]^{-1},\\
{\M}_4&=&
 \mbox{tr}\!\left\{\a_2(\pc-\q_2+m_c)\a_1(-\pb-\q_3+m_c)\a_3\;{\P}_S\right\}\n
&\times&
 \left[k_2^2-2(p_{c}k_2)\right]^{-1}\left[k_3^2+2(p_{\bar{c}}k_3)\right]^{-1},\\
{\M}_5&=&
 \mbox{tr}\!\left\{\a_2(\pc-\q_2+m_c)\a_3(-\pb+\q_1+m_c)\a_1\;{\P}_S\right\}\n
&\times&
 \left[k_2^2-2(p_{c}k_2)\right]^{-1}\left[k_1^2-2(p_{\bar{c}}k_1)\right]^{-1},\\
{\M}_6&=&
 \mbox{tr}\!\left\{\a_3(\pc+\q_3+m_c)\a_2(-\pb+\q_1+m_c)\a_1\;{\P}_S\right\}\n
&\times&
 \left[k_3^2+2(p_{c}k_3)\right]^{-1}\left[k_1^2-2(p_{\bar{c}}k_1)\right]^{-1},\\
{\M}&=&{\M}_1+{\M}_2+{\M}_3+{\M}_4+{\M}_5+{\M}_6,
\end{eqnarray}
with the property ${\M}_1={\M}_6$, ${\M}_2={\M}_5$, ${\M}_3={\M}_4$.
The color factor is universal and is equal to $d^{abc}/4\sqrt{3}$.
This set of diagrams is complete; no other diagrams can contribute at the order
${\O}(\alpha_s^3)$ to the production of a meson with the given quantum numbers
$J^{PC}=1^{+-}$.

The amplitudes ${\M}_i$ contain spin projection operators which discriminate
the spin-singlet and spin-triplet $c\bar{c}$ states:
\begin{eqnarray}
{\P}_{S{=}0} &=& (\pb-m_c)\gamma_5(\pc+m_c)\cdot (2m_c)^{-3/2},\\
{\P}_{S{=}1} &=& (\pb-m_c)\a_{\psi}(\pc+m_c)\cdot (2m_c)^{-3/2},
\end{eqnarray}
where $m_c$ is the charmed quark mass.
These projectors are orthogonal to each other, as they should be: 
$tr\{{\P}_0 \overline{{\P}_1}\}=0$. For the $^1P_1^{[1]}$ state
we evidently have to use the projector ${\P}_0$.

The orbital angular momentum $L$ is associated with the relative momentum $q$
of the quarks in a bound state. The relative momentum $q$ is defined as
\begin{equation}
p_c=\frac{1}{2}p+q,\quad p_{\bar{c}}=\frac{1}{2}p-q.
\end{equation}
According to a general formalism developed in \cite{Guber,Krase}, the terms
showing no dependence on $q$ are identified with the contributions to the $L=0$
state; the terms linear in $q^\alpha$ are related to the $L=1$ state with the
proper polarization vector $\epsilon^\alpha$ (see below); the quadratic terms
$q^\alpha q^\beta$ refer to the $L=2$ state with the polarization tensor
$\epsilon^{\alpha\beta}$; and so on.
The decomposition of ${\M}$ in powers of $q$ is carried out by expanding the 
subprocess amplitude as
\begin{equation}\label{Taylor}
{\M}(q)={\M}|_{q{=}0} + q^\alpha(\partial{\M}/\partial q^\alpha)|_{q{=}0} + ...,
\end{equation}
where $q$ is assumed to be a small quantity. The amplitude ${\M}(q)$ has
to be multiplied by the bound state wave finction $\Psi(q)$ and integrated
over $q$. A term-by-term integration of Eq.(\ref{Taylor}) is performed
using the relations
\begin{eqnarray}
\int \frac{d^3q}{(2\pi)^3}\Psi(q)&=&\frac{1}{\sqrt{4\pi}}{\R}(x{=}0),\\
\int \frac{d^3q}{(2\pi)^3}q^\alpha\Psi(q)&=&
    -i\epsilon^\alpha\frac{\sqrt{3}}{\sqrt{4\pi}}{\R}'(x{=}0),
\end{eqnarray}
etc., where ${\R}(x)$ is the radial wave function in the coordinate 
representation (the Fourier transform of $\Psi(q)$).
This formula completes our derivation of the production matrix element.
The resulting expression has been explicitly tested for gauge invariance by 
substituting the gluon momentum $k_i$ for the polarization vector $\epsilon_i$.
We have observed gauge invariance even with off-shell initial gluons.


\begin{center}
\begin{table*}[t]\label{Tab_etac}
\caption{Sets of NME's for \e production as determined from the different fits}
\begin{tabular}{lcccc}
\hline \hline
  & JH set 1 & JH set 2 &~Kniehl \et ~\cite{BK}~&~Gong \et ~\cite{GW}\\[1mm]
\hline
$\left\langle \O^{\eta_c}\bigl[^1S_0^{[1]}\bigr]\right\rangle/\mbox{GeV}^3$
  & 0.39    &  0.39     & 0.44               &  0.39  \\
$\left\langle \O^{\eta_c}\bigl[^3S_1^{[8]}\bigr]\right\rangle/\mbox{GeV}^3$
  &  0.0    &  0.0      & 0.304              &  0.097 \\
$\left\langle \O^{\eta_c}\bigl[^1S_0^{[8]}\bigr]\right\rangle/\mbox{GeV}^3$
  & $(1.4 \pm 0.3) \cdot 10^{-4}$ & $(5.3 \pm 0.7) \cdot 10^{-4}$~&~0.00056~&~-0.0015\\
$\left\langle \O^{\eta_c}\bigl[^1P_1^{[8]}\bigr]\right\rangle/\mbox{GeV}^5$
  & $0.069 \pm 0.006$   & $0.072 \pm 0.006$  &-0.02724 & -0.0642\\
\hline \hline
\end{tabular}
\end{table*}
\end{center}

\acknowledgments

We would like to thank H. Jung for his interest, very useful discussions and 
important remarks. This work was supported by the DESY Directorate in the 
framework of Moscow-DESY project on Monte Carlo implementations for HERA-LHC.

\end{document}